\documentstyle[12pt,aaspp4]{article}              

\lefthead{}
\righthead{}
\slugcomment{PASP, accepted}
\received{}
\input epsf.tex

\begin{document}

\title{Biases in Expansion Distances of Novae Arising from the Prolate
Geometry of Nova Shells}

\author{Richard A.\ Wade\altaffilmark{1,2}} 

\author{Jason J.\ B.\ Harlow\altaffilmark{1,3}}

\and 

\author{Robin Ciardullo\altaffilmark{1,4}}

\altaffiltext{1}{Department of Astronomy \& Astrophysics, The Pennsylvania 
State University,\\ 525 Davey Laboratory, University Park, PA 16802-6305}
\altaffiltext{2}{wade@astro.psu.edu}
\altaffiltext{3}{Present address: Department of Physics, University
of the Pacific, 3601 Pacific Ave., Stockton, CA 95211-0197.  
email: jharlow@uop.edu} 
\altaffiltext{4}{rbc@astro.psu.edu}

\begin{abstract} 
Expansion distances (or expansion parallaxes) for classical novae are
based on comparing a measurement of the shell expansion velocity,
multiplied by the time since outburst, with some measure of the
angular size of the shell.  We review and formalize this method in the
case of prolate spheroidal shells.  For such shells there is no unique
angular size except when the shell is seen pole-on, and several
different measures of angular size have been used in the literature.
We present expressions for the maximum line-of-sight velocity from a
complete, expanding shell and for its projected major and minor axes,
in terms of the intrinsic axis ratio and the inclination of the polar
axis to the line of sight.  For six distinct definitions of ``angular
size'', we tabulate the error in distance that is introduced under the
assumption of spherical symmetry (i.e., without correcting for
inclination and axis ratio).  The errors can be significant and
systematic, and can affect studies of novae whether considered
individually or statistically.  Each of the six estimators
overpredicts the distance when the polar axis is close to the line of
sight, and most underpredict the distance when the polar axis is close
to the plane of the sky. Use of the straight mean of the projected
semimajor and semiminor axes gives the least distance bias for an
ensemble of randomly oriented prolate shells, and we recommend this
method when individual inclinations and axis ratios cannot be
ascertained.  The best individual expansion distances, however, result
from a full spatio-kinematic modeling of the nova shell.  We discuss
several practical complications that affect expansion distance
measurements of real nova shells.  We recommend that nova shell
expansion distances be based on velocity and angular size measurements
made contemporaneously if possible, and that the same ions and
transitions be used for the imaging and velocity measurements.  We
emphasize the need for complete and explicit reporting of measurement
procedures and results, regardless of the specific method used.

\end{abstract}

\keywords{novae --- circumstellar matter}


\section{INTRODUCTION}\label{intro} 

The distance to a classical nova in the Galaxy is best inferred by
comparing the angular size of the resolved nova shell with the size
calculated from its rate of expansion and the time since the shell was
ejected.  However, if nova shells are ellipsoids of revolution
(spheroids) rather than spherical, then the concept of ``angular
size'' is ambiguous, and the expansion velocity along the line of
sight does not correspond to the transverse expansion velocity.  Thus
the use of formulas that are valid in the spherical case will lead to
erroneous distance estimates. Individual distance estimates may be too
large or too small, depending on the true axis ratio of the nova shell
and its inclination to the line of sight.  Furthermore, these errors
do not necessarily average toward zero when ensemble averages are
taken.  In this paper, we consider systematic errors in estimates of
nova shell expansion distances, and recommend procedures to minimize
the errors.

\subsection{The Usefulness of Nova Distances}

As with all classes of astronomical objects, our understanding of the
classical nova phenomenon depends on having accurate estimates of the
distances of these objects.  In turn, having a well-founded
understanding of the distances and luminosities of novae allows them
to be studied as astrophysical objects and to be exploited for other
purposes.

The need for accurate distances is evident, both for novae taken
individually and for novae used collectively, i.e., in a statistical
fashion.  An accurate distance to an individual nova, combined with
good coverage of the outburst light curve and a knowledge of the
interstellar extinction, can allow the theory of the nova outburst to
be verified and further developed.  For example, it is possible to
check whether there is a phase after maximum light during which the
nova's luminosity is close to the Eddington limit.  All inferences
about the mass of the shell that is ejected during the outburst depend
on some power of the distance, through establishing the volume
occupied by the emitted gas.  At late stages in the evolution of a
nova, when the shell can be resolved from the central binary star, the
distance is needed to convert the angular size of the shell into a
linear size, so that the physical conditions in the ejected gas
(ionization, excitation) can be related to the ionizing flux from the
central white dwarf (the post-nova) and its accretion disk.  An
accurate distance to a classical nova also allows the modeling of the
accretion process in the post-nova system; without a distance
constraint (leading to a constraint on the luminosity), 
it has proven impossible to infer uniquely the mass
accretion rate onto the white dwarf (Wade 1988, Wade \& Hubeny 1998).

Treated collectively, novae have the possibility of providing a
secondary or even primary distance indicator for the extragalactic
distance scale.  Since novae are present in galaxies of all Hubble
types, they have the potential to be used directly to compare and
unite the distance scales of spiral and elliptical galaxies.  The
so-called maximum magnitude -- rate of decline (MMRD) relation gives
the visual absolute magnitude at maximum light, from a measurement of
the rate of decline after maximum (or equivalently the time taken to
decline 2 or 3 magnitudes).  The shape of the mean MMRD curve and the
dispersion around this mean relation have been found from observations
in external galaxies (e.g., Della Valle \& Livio 1995). For the MMRD
relation to be a primary distance indicator, however, the zero-point
calibration must be provided by Galactic novae.  Another proposed
distance indicator is the absolute magnitude at 15 days past maximum
light ($M_{15}$), where the dispersion in absolute magnitude is small
for all novae taken without regard to speed class.  The same remarks
about zero-point calibration apply to this method.\footnote{As one
step in calibrating the MMRD relation, Cohen (1985) adjusted $M_V$ at
maximum for her best observed novae, assuming that they had {\em
identical} $M_{15}$; this step should be replaced by actually
measuring the dispersion in $M_{15}$.}

Even the nearest Galactic novae are generally too distant for direct
trigonometric parallax measurements.  Instead, indirect methods of
distance estimation are often used, based on the Galactic rotation
curve, the total amount of interstellar reddening, the presence or
absence of discrete components (``clouds'') in interstellar absorption
lines, etc.  The only geometrical (hence fundamental) method is that
of ``expansion distance'' (also referred to as ``expansion
parallax''), in which the measured angular size of the resolved nova
shell is compared with the linear size of the shell; the latter is
calculated from the expansion speed of the shell gas and the known
time since the outburst.  In her work on the MMRD relation, Cohen
(1985) had only eleven novae with well-observed expansion distances
and suitable coverage of their light curves.  This is largely because
the surface brightness of a nova shell declines rapidly with time
since outburst, so that by the time the shell is large enough to be
resolved from the ground, it is often too faint to observe.  Since
Cohen's study, several additional novae have had good light curve
coverage, and expansion distances may become available for these from
ground-based observations and especially from the {\it Hubble Space
Telescope\/}\footnote{Narrow-band imagery of recent novae has
been carried out with the HST Wide Field/Planetary Camera 2, used in
``snapshot'' mode, as part of program 7386; these images are public.
HST imagery of somewhat older nova shells is discussed in Gill \&
O'Brien (2000).}, since the latter can resolve some shells within a
few months or years of the outburst.

The simplest way to derive an expansion distance is to assume (often
implicitly) that the nova shell is expanding spherically
symmetrically, hence that the transverse velocity of gas in the plane
of the sky is the same as the radial velocity of gas moving directly
along the line of sight.  What if the ejection of the shell is
asymmetric?  To be specific, suppose the shell expands as a spheroid,
the simplest generalization from the spherically symmetric case, and
one that suffices to describe many actual nova shells.
First, the projected image of the nova shell will not be circular for
most orientations, and thus there will be an ambiguity in what is
meant by angular size.  Second, the maximum velocity along the line of
sight will usually not correspond to {\em either} the ``polar'' or
``equatorial'' expansion velocity.  For example, suppose that the
angular size is taken to be the largest projected ``radius'' of the
nova shell, which is perhaps the easiest size parameter to estimate on
a barely resolved image.  If all nova shells were oblate, then the
calculated expansion distance (based on the assumption of spherical
symmetry) would always be less than the true distance, because the
line-of-sight velocity would be smaller than the transverse expansion
velocity.  The resulting nova distance scale would be too short.  On
the other hand, if all nova shells are prolate, than the distance to
an individual object may be underestimated or overestimated, depending
on the orientation and the ratio of major and minor axes.  While it is
clear that the distance to an individual nova can be in error as a
result, it was not made clear until the work of Ford \& Ciardullo
(1988; hereinafter FC88) that in the prolate case, a systematic error
might remain, even after averaging over an ensemble of novae that are
taken to be randomly oriented in space.  In their analysis, FC88 made
the assumption stated above as an example, that the angular size of
the nova shell is taken to be the major axis of the projected
image. However, all workers do not make this identification.  For
example, Cohen \& Rosenthal (1983) did use the projected semimajor
axis for the angular size, but Cohen (1985) used an angle-averaged
radius.  What way is best? A goal of this paper is to extend the FC88
analysis to include consideration of six distinct yet plausible ways
of defining the angular size of the shell.

As the number of Galactic novae with well observed light curves
increases, and with the much greater resolving power provided by
adaptive optics and HST, it is likely that the calibration of the MMRD
and $M_{15}$ relations will be improved, but the question of possible
systematic errors in the distances becomes more important.  This is
especially so, if shell morphology is related to nova speed class, as
has been suggested by Slavin, O'Brien, \& Dunlop (1995). Likewise, as
more expansion distances for individual novae become available, it is
important to have clearly in mind whether and how much these distances
may be in error, as the result of measuring uncertainties and modeling
assumptions.

\subsection{Prolate or Oblate?}

Theoretical arguments have been made favoring both oblate and prolate
geometries for nova shells (Porter, O'Brien, \& Bode 1998 and
references therein).  Empirically, it is now the consensus that,
to the extent that nova shells can be described by spheroids, they
are either prolate or spherical, but not oblate.  FC88 discussed the
few cases of resolved nova shells known at the time, in terms of
whether they were elongated along one axis (prolate spheroids) or
compressed along one axis (oblate spheroids).  FC88 noted that most
data on the shapes of nova shells were consistent with spherical or
prolate geometries, but categorized the shell of nova HR Del
1967 as oblate. For this object, early models by Hutchings (1972) and
Soderblom (1976) indeed suggested an oblate symmetry.  A spatio-kinematic
model by Solf (1983), however, has clearly shown that the resolved
shell of HR Del is consistent with a prolate geometry, and not
consistent with being oblate.  Slavin, O'Brien, \& Dunlop (1995)
carried out imaging of nova shells using narrow band filters; in
particular they have obtained images at several different tilts of an
interference filter with nominal wavelength 6560 \AA\ (17 \AA\ FWHM),
which allowed them to distinguish crudely between gas approaching or
receding from the observer.  Their data are clearly consistent with
the shells being prolate, not oblate, if they depart detectably from
spherical symmetry.  Other spatio-kinematic studies, for example of
the shell around nova DQ Her 1934 (e.g. Herbig \& Smak 1992) also
indicate prolate symmetry.  Therefore we proceed with the
assumption that to first approximation, nova shells are prolate
spheroidal shells, with their properties in projection specified by
their axis ratio and the inclination of the polar axis to the line of
sight.

In Section 2 of this paper, we investigate several different ways of
defining the angular size of the resolved nova shell.  We derive the
projected size and shape of the shell and the maximum radial velocity
of gas in the shell, as functions of the intrinsic axis ratio and the
inclination of the polar axis.  We then derive analytic expressions
that give the inferred distance in terms of the true distance, as
a function of axis ratio and inclination.  We tabulate results for a
variety of cases, using six definitions of ``angular size.''  In Section 3
we investigate how these various definitions of expansion distance
behave, both for individual objects and when averaged over an ensemble
of nova shells oriented randomly in space.  We also discuss some
practical matters relating to the measurement of the expansion speed
and angular size of nova shells. We summarize our findings in Section 4.


\section{EXPANSION DISTANCE ESTIMATORS FOR PROLATE SPHEROIDAL NOVA SHELLS}

When viewed at an inclination angle $i$, a prolate spheroid at
distance $d$ and with principal axis ratio $b/a \le 1$ will appear
projected on the sky as an ellipse with apparent axis ratio $b_\perp /
a_\perp \le 1$.  We define several quantities, which appear repeatedly
in the discussion to follow:
$$f_1 = \sqrt{1-e^2\sin^2 i} $$
$$f_2 = \sqrt{1-e^2\cos^2 i} $$
$$f_3 = \sqrt{1-e^2} = b/a $$
The auxiliary quantity $e$ is the ``eccentricity'' of the prolate spheroid,
in the sense that $b^2 = a^2(1-e^2)$ relates the major
and minor axes, $a$ and $b$ respectively, of an ellipse.

We have the following relations between the axes of the spheroid in
space, $a$ and $b$, and the (linear) principal axes of the projected ellipse,
$a_\perp$ and $b_\perp$ (see Appendix A).
$$a_\perp = f_2 a $$
$$b_\perp = b  = f_3 a = (f_3/f_2)a_\perp$$
Let $v_0$ denote the expansion speed along the major (polar) axis
of the spheroid. Then from Appendix B the maximum projected (line-of-sight) speed is
$$v_{\rm max} = f_1 v_0 = f_1 (a/t),$$ 
where $a$ is the semimajor axis of the spheroid when the age of the
nova remnant is $t$.  (Constant expansion speed is assumed.)  Also, a
distance $x$ measured in the plane of the sky corresponds to an angle
$\rho_x = x / d$ (radians), where $d$ is the true distance of the
nova.

The essence of the expansion distance method is to compare an estimate
of the linear size of the nova shell, $vt$, with an estimate of the
angular size, $\rho$.  The estimator formula 
\begin{equation}
\hat{d} = {v_{\rm max} t \over \rho},
\end{equation}
where $\rho$ is any angular radius, recovers the true distance $d$ in
the case of a spherically symmetric expanding shell.  This is because
$v_{\rm max} = v_0$ (by symmetry) and all angular radii are equal to
$a /d$ where $a$ is the true linear size of the shell at time $t$:
\begin{equation}
\hat{d} = {v_{\rm max} t \over \rho} = {v_0 t \over (a/d)} = d.
\end{equation}

For a prolate spheroid, in general $v_{\rm max} \ne v_0$, and there is
no unique measure of the angular size $\rho$.  Given independent
knowledge of $i$, the apparent ratio $b_\perp / a_\perp = f_3 / f_2 =
(1-e^2)^{1/2} / (1-e^2\cos^2 i)^{1/2}$ can be inverted to find $e$.
In this case equation (1) can be used, with corrections to convert the
measured speed $v_{\rm max}$ into $v_0$ and the apparent semimajor
axis $\rho_1 = a_\perp / d$ into $a/d$, to recover the true distance.
In symbols,
\begin{equation}
\hat{d} = {(v_{\rm max}/f_1) t \over (\rho_1/f_2)} = {v_0 t \over (a/d)} = d.
\end{equation}

In general, however, the inclination of the spheroid in space is not
known, so the correction factors are not known, but the simple formula
$\hat{d} = {v_{\rm max} t / \rho_1}$ does not recover $d$.
Furthermore, there is no reason any longer to define $\rho$ as the
apparent semimajor axis --- the apparent minor axis or some average
indicator of the apparent size of the nova shell could be used
instead.  

We consider six possible definitions of $\rho$, and for each we
investigate how large an error is made in estimating $d$ using
equation (1).  This question is addressed both for individual novae,
with particular values of $b/a$ and $i$, and for statistical ensembles
of nova shells, where averages are taken over random orientations for
a fixed value of $b/a$.

The six choices for $\rho$ are:
$$\begin{array}{lclcl}
\rho_1 & = &a_\perp / d & = & f_2\times a/d\\
\rho_2 & = &b_\perp / d & = & f_3\times a/d\\
\rho_3 & = &(a_\perp + b_\perp) / 2d & = & (f_3 + f_2)/ 2\times a/d\\
\rho_4 & = &\sqrt{a_\perp b_\perp}/ d & = & \sqrt{f_3f_2}\times a/d\\
\rho_5 & = & 2 b_\perp K(k) / \pi d & = & 2 f_3 K(k) / \pi\times a/d\\
\rho_6 & = &2(a_\perp^{-1}+b_\perp^{-1})^{-1}/d &=& 2(f_2^{-1}+f_3^{-1})^{-1}\times a/d
\end{array}$$

The first two choices are the apparent major and minor axes of the
projected ellipse.  The third choice is the arithmetic mean of
$\rho_1$ and $\rho_2$.  The fourth choice is the geometric mean of
these.  The fifth definition, $\rho_5$, is the angle-averaged apparent
``radius'' of the shell. Here $K(k)$ is the complete elliptic integral
of the first kind (see Appendix C).  The argument $k$ is given by
$k^2 = 1 - (k^\prime)^2 = 1 - (b_\perp / a_\perp)^2 = 1 - (f_3/f_2)^2.$
The sixth choice is the harmonic mean of $\rho_1$ and $\rho_2$.

Corresponding to each $\rho_i$ is a distance estimator $\hat{d}_i =
v_{\rm max} t / \rho_i$.  Given in Table 1 are 
$k^\prime = b_\perp/a_\perp$, $v_{\rm max}/v_0$, $a_\perp/a$, and
$\hat{d}_j/d, (j=1,2,\dots 5)$ for several combinations of true axis
ratio $b/a$ and inclination $i$.  Since $\hat{d}_6/d = (\hat{d}_1 +
\hat{d}_2)/2d$, it is not shown separately in Table 1.

The probability that the polar axis of a randomly oriented
spheroid makes an angle between $i$ and $i+di$ with the line of sight 
is $P(i)di = (\sin i) di$.  Each of the estimates $\hat{d}_j$ can
be averaged over angle with $P(i)$ as the weighting function:
\begin{equation}
\langle \hat{d}_j\rangle = \int_0^{\pi/2}P(i)\hat{d}_j(i) di.
\end{equation}
The $P(i)$--weighted average values of $\hat{d}_j/d$ were computed
numerically and are shown in Table 1.  For example,
$\langle\hat{d}_1\rangle/d = 0.937$ for $b/a = 0.80$.


\section{RESULTS AND DISCUSSION}

\subsection{The Typical Size of Errors in Expansion Distances}

Inspection of Table 1 reveals several features of the various distance
estimators $\hat{d}_i$.  Every estimator gives the correct distance
for spherically symmetric nova shells, as expected.  For prolate nova
shells, every estimator overpredicts the distance when the polar axis
is close to the line of sight, because $v_{max}$ does not correspond
to the projected angular size.  In this case, all estimators are
equally poor, and increasingly so as the axis ratio $b/a$ becomes more
extreme.  All estimators except $\hat{d}_2$ underpredict the distance
for prolate nova shells when the polar axis is close to the plane of
the sky.  In this case, the error increases as the axis ratio
decreases, but $\hat{d}_5$ and $\hat{d}_6$ are better than the other
estimators.

Finally, when considering an ensemble average of shells with random
orientations, the best average distance is produced using $\hat{d}_3$,
which is based on the straight mean of the projected semimajor and
semiminor axes.  Since this is true for each intrinsic axis ratio
considered separately, it will also be true for an ensemble of
randomly oriented nova shells that has a mixture of axis ratios.

Observations of resolved nova shells show a variety of projected axis
ratios.  For example, nova DQ Her 1934 has a projected axis ratio of
$k^\prime = b_\perp/a_\perp = 0.73$ (Herbig \& Smak 1992), while nova
FH Ser 1970 has $k^\prime = 0.91$ (Slavin, O'Brien, \& Dunlop 1995).
Nova HR Del 1967 has $k^\prime = 0.56$ according to Solf (1983).
Slavin et al.\ (1995) find $k^\prime = 0.75$ for HR Del viewed in
H$\alpha$, but a more elongated image (same projected major axis,
shorter projected minor axis) in the light of
[\ion{O}{3}].\footnote{Slavin, O'Brien, \& Dunlop (1994) discuss the
different appearance of HR Del in different ions as perhaps arising
from density or composition differences of the gas ejected in
different directions from the central star. Another possibility is a
difference in ionizing flux between the poles and the equator of the
shell, with the equator being shielded by the accretion disk around
the star (cf.\ the discussion of nova FH Ser 1970 by Gill \& O'Brien
2000).}  Solf finds the intrinsic axis ratio for HR Del to be $b/a =
1/3$, possibly the most extreme of the well-studied shells.  Thus
Table 1 covers most of the axis ratios observed to date.  If there
were a shell with $b/a = 0.4$ viewed pole-on, the expansion distance
method would yield an estimated distance a factor of 2.5 larger than
the true distance, regardless of the specific estimator used.  The
median inclination of the polar axis is 60$^\circ$, and 80 per cent of
all randomly oriented nova shells should have inclinations between
26$^\circ$ and 84$^\circ$.  Thus {\em typical} individual errors in
nova expansion distances will be of order tens of percent, unless the
inclination and axis ratio are known.  Barring the tell-tale evidence
of high inclination, given by an eclipsing binary system, the only way
to derive the inclination of a nova shell is to construct a so-called
spatio-kinematic model, in which spectra from multiple positions
across the resolved shell are simultaneously modeled (e.g., HR Del,
Solf 1983; DQ Her, Herbig \& Smak 1992, Gill \& O'Brien 2000).

\subsection{Practicalities of Measuring Nova Shells}

The nova shell literature contains a mixture of techniques for
measuring the expansion velocities and angular sizes.  Given that nova
shells typically are bright enough to detect only for a few decades
after outburst and hence have small angular sizes if they are
resolved, this is not surprising.  Nevertheless, it is clear that
velocities and angular sizes have not always been combined in a
self-consistent fashion, even in cases where the inclination and axis
ratio are known.  Herbig \& Smak (1992) discuss this point at length
for the case of nova DQ Her 1934, comparing several different
spatio-kinematic studies that arrived at widely differing distance
estimates.  For DQ Her the issue is that the shell has a finite
thickness; it is therefore possible erroneously to combine a velocity
measured from the extreme outer edge with, say, a size measured from
the ridge line of an image, which represents a position {\em within}
the shell.  Martin (1989) also discusses this point. To avoid
confusion, it is highly desirable for observers reporting either a
line-of-sight expansion velocity or an angular size to be explicit
about exactly what was measured.  We suggest that angular size
measurements be referred to a contour level that encloses a stated
percentage of the nova shell flux. (Care needs to be taken, if imaging
is done through a narrow-band filter that excludes part of the
shell emission.)  Likewise, emission line velocities may be
measured at a level above the continuum such that a stated percentage
of the total line flux is at less extreme velocities.

If image deconvolution methods (e.g., the MEM and CLEAN algorithms)
are used to ``sharpen'' the image, special care needs to be taken in
determining and reporting any effect this has on measurements of the
``edge'' of a nova shell, especially if the method does not conserve
flux.

In addition to avoiding possible ambiguities caused by the thickness
of the shell, authors need to be clear about what sort of angular size
is being reported.  Cohen \& Rosenthal (1983) used $\rho_1$, the
projected semimajor axis of the shell, and FC88 modeled the systematic
errors in nova expansion distances assuming $\rho_1$, as it is perhaps
the easiest to measure for barely resolved shells.  Cohen (1985),
however, derived angular sizes by carefully modeling the appearance of
a (spherically symmetric) shell superimposed on a central star, taking
into account the point spread function of the optical system. This
technique gives $\rho_5$.  Others, e.g., Shin et al.\ (1998), have
been less careful, merely measuring the FWHM of the nova image and
subtracting a stellar FWHM in quadrature to obtain a
``characteristic'' size. While this method is adequate to demonstrate
angular extension of the shell, it is essentially useless for distance
estimation, since it wrongly assumes that the profile of a
seeing-convolved image of a shell is gaussian, and it does not take
into account light from the central star.

Additional complications arise if the projected image of the nova
shell is elongated but not strictly elliptical, or if the outline of
the shell is incomplete.  Herbig \& Smak (1992) demonstrated that the
``equator'' of the DQ Her shell is constricted in both angular extent
and velocity, and were careful to distinguish the measured equatorial
velocity from the modeled minor-axis velocity of the spheroid.  If
there is emission from gas {\em beyond} the main elliptical outline,
as reported for DQ Her (Slavin, O'Brien, \& Dunlop 1995) and RR Pic
(Gill \& O'Brien 1998), care must be taken that the extreme velocity
used to estimate distance corresponds to the shell from which the
angular size was taken, and not the extended halo.  Furthermore, if
the spheroidal shell is incomplete, consisting only of an equatorial
``belt'' and polar ``blobs'', then for most inclination angles, there
will be no emitting gas at the internal angle $\theta$ (see Figure 2)
that would correspond to $v_{max}$ in a filled shell.  Likewise, there
may not be any gas emitting at the tangent to the line of sight, which
defines the projected major axis for a complete shell (Appendix A).
In such a case, our prescription for distance estimators formally
breaks down.  However, FC88 give a partial discussion of this case in
the limit of a narrow equatorial belt and small polar caps, giving
expressions for $\hat{d}_1/d$ and its average for random
orientations. FC88 find that individual nova distances for the ``belt
and blobs'' case can be over- or underestimated, up to the same
extremes as for the complete shell case, and the angle-averaged
distance is underestimated by an amount very similar to the complete
shell case discussed here.

Another complication arises from the fact that some nova shells have
been observed to change their projected shape in the few years
immediately after outburst.  A notable case is that of nova V1974 Cyg
1992, which was observed with long-baseline interferometry at radio
frequencies (Hjellming 1996).  The radio data suggest a model in which
the outer and inner faces of the expanding shell have different
(triaxial)  ellipsoidal representations.  Moreover, careful models
constructed to account for radio observations of novae (Hjellming et
al.\ 1979; Seaquist et al.\ 1980) demonstrate that the entire
expanding gas cloud does not contribute equally to the optical
emission line radiation, since the shell is likely more dense at the
inner boundary. The emissivities of the various ions and the fraction
of the expanding shell that contributes to the flux both change with
time, as the nova shell expands and the post-nova central binary
(which photoionizes the shell) recovers from the eruption.  The size
and morphology of a given nova shell image may differ, depending on
which transition is used to image it.  As discussed in Section 3.1, HR
Del differs considerably in appearance, depending on whether the shell
is imaged in $\lambda$5007 [\ion{O}{3}] or in
$\lambda\lambda$6548/6563/6584 H$\alpha$+[\ion{N}{2}].  For all these
reasons, it is best to measure the expansion velocity and the angular
size of a shell contemporaneously if possible; this will assure that the same
parcels of gas are used for both measurements.

Finally, since the central star of a nova system is a cataclysmic
binary, it is possible to conflate an emission line from an accretion
disk or magnetically channeled accretion flow in the binary with the
same line (or a nearby one) from gas in the nova shell.  Since orbital
speeds of gas in the inner accretion disk may be of order
$10^3$~km~s$^{-1}$, there is the possibility to mis-measure the
expansion velocity of the shell, unless care is taken to measure
$v_{max}$ from a long-slit spectrum oriented along the projected major
axis.  (Recall from Appendix B that the point of the projected image
at which $v_{max}$ is attained is generally not the same point as the
central star.)

These examples of traps and complications in the estimation of nova
expansion distances are intended to reinforce a plea that observers
give a full description of their methods and results.  Depending on
the particular nova, the information available, and the goals of the
study, one method of proceeding or another may be appropriate ---
perhaps there is no single way to estimate an expansion distance that
is best in every instance.  To allow results from different studies to
be combined correctly, the need for careful documentation is evident.


\section{SUMMARY AND CONCLUSION}

We have reviewed and formalized the method of nova shell expansion
distances as a means of estimating the distances to classical novae.
This method combines a measurement of the shell expansion velocity
(multiplied by the time since outburst) with some measure of the
angular size.  Expansion distances for novae underlie the calibration
of the MMRD and $M_{15}$ relations and also form the basis for
astrophysical studies of individual novae and their remnants. It is
therefore important to adopt methods of measurement that minimize any
possible bias in the distances that results from incomplete
information about the shape or orientation of the nova shells.

Many resolved shells exhibit significant prolate symmetry, so that
there is no unique angular size except when the shell is seen pole-on.
We developed analytic expressions for the maximum line-of-sight
velocity from a complete, expanding prolate spheroidal shell and for
its projected major and minor axes, as functions of the intrinsic axis
ratio and the inclination of the polar axis to the line of sight.  For
six definitions of ``angular size'', we then computed the error
introduced by deriving a distance using the assumption of spherical
symmetry (i.e., without correcting for inclination and axis ratio).
The errors can be significant and possibly systematic, affecting
studies of novae whether considered individually or statistically.

The definition of angular size that results in the least errors at the
extremes is $\rho_6$, the harmonic mean of the projected semimajor and
semiminor axes.  However, the definition that results in the least
bias when an ensemble of randomly oriented prolate shells is
considered is $\rho_3$, the straight mean of the projected semimajor
and semiminor axes, and we recommend this method when individual
inclinations and axis ratios cannot be ascertained.  The
$\rho_3$--based method is always as good or better than than the
$\rho_1$ method (projected semimajor axis alone).  The best individual
expansion distances result from a full spatio-kinematic modeling of
the nova shell, using spectroscopy of emission lines at multiple
locations across the resolved shell.

We have discussed practical issues and made recommendations for observers
who make measurements of either the maximum line-of-sight velocity and the
angular size of a resolved nova shell.  The velocity measurement may
be complicated by the presence of line emission from the central
cataclysmic binary star, and if the spheroidal shell is not complete,
the theoretical maximum velocity may not be observed at all.  The
correct application of angular size measurements can be compromised by
convolution with the image point spread function, by improper
technique, or by incomplete reporting.  For best results, velocity and
angular size measurements should be made contemporaneously, and must
refer to the same features of the shell.  Observers are encouraged to
report as completely as possible the measurements they have made.
Estimates of nova distances by the shell expansion method (or any
other method) should be accompanied by a discussion of both random and
systematic errors, including possible effects due to unaccounted-for
departures from spherical symmetry, as discussed in this paper.


\acknowledgements

Support for this work was provided by NASA through grant number
GO-07386.01 from the Space Telescope Science Institute, which is
operated by AURA, Inc. under NASA contract NAS 5-26555.  This research
has made use of the Simbad database, operated at CDS, Strasbourg,
France.


\newpage

\appendix

\section{THE PRINCIPAL AXES OF THE PROJECTED ELLIPSE}

A prolate spheroidal shell centered on the origin, with its major axis aligned
with the $z$ axis, is described by the equation:
$$ \frac{x^2}{b^2} + \frac{y^2}{b^2} + \frac{z^2}{a^2} = 1 $$ 
with $b<a$. The eccentricity of the ellipse, $e$, is defined by $b^2 =
a^2(1-e^2)$.

The observer's line of sight, taken to be in the $xz$ plane, makes
an angle $i$ with the $z$ axis (polar axis).  This observer sees a projected
ellipse with semimajor axis $a_\perp$ and semiminor axis $b_\perp$.
The prolate symmetry around the $z$ axis gives the result:
$$ b_\perp=b=a \sqrt{1-e^2}.$$

The semimajor projected axis $a_\perp$ can be found using the geometry
shown in Figure 1. The intersection of the spheroidal surface and the
$xz$ plane is an ellipse described by
$$ \frac{x^2}{b^2} + \frac{z^2}{a^2} = 1,$$
or
$$ x^2 = a^2(1-e^2) - z^2(1-e^2).$$

The line of sight is described generally by
$$ z = c + x\cot i.$$
The {\em tangent} line of sight passes through point $A$, and for this
line $c$ is defined by the condition that the line intersects the
ellipse exactly once.  Using the equation of the line to substitute
for $z$ in the equation of the ellipse, it is seen that
$$ x^2[1 + (1-e^2)\cot^2 i] + x[2 c (1-e^2)\cot i] + (c^2 - a^2)(1-e^2)  = 0 $$

This equation, quadratic in $x$, has a single solution (tangent
condition) only when the discriminant, D, is equal to zero:
$$D = 4c^2 (1-e^2)^2 \cot^2 i - 4[1 + (1-e^2)\cot^2 i](1-e^2)(c^2 - a^2) = 0.$$
The $z$-intercept of the line is thus
$$ c = a \sqrt{1 + (1-e^2)\cot^2 i},$$
and the projected semimajor axis is
$$ a_\perp = c\sin i = a\sin i \sqrt{1+(1-e^2)\cot^2 i} = 
\sqrt{a^2\sin^2 i + b^2 \cos^2 i} = a\sqrt{1 - e^2 \cos^2 i}.$$
It is easy to see that $a_\perp > b_\perp$.

The tangent method for finding the projected ellipse was used as long
ago as Hubble (1926), although he used it only for oblate spheroids
and measured $i$ from the equator rather than the pole of the
spheroid.

\section{THE MAXIMUM LINE-OF-SIGHT VELOCITY}

As before, let the first quadrant of the $xz$ plane contain the
observer's line of sight, in a direction defined by the unit vector
$\hat{n} = (\sin i, \cos i)$ where $i$ is the angle between the major
axis of the spheroid and the observer.  By symmetry the maximum
projected (line-of-sight) velocity of the ellipsoid will be associated
with a point that lies in the $xz$ plane, and it
suffices to consider the plane ellipse
$$ \frac{x^2}{b^2} + \frac{z^2}{a^2} = 1,$$
or $ z^2 = a^2 - x^2(1-e^2)^{-1}$ where $b^2 = a^2(1-e^2)$ as before. 
Let $\theta$ by the polar angle defined by $ x = z \tan \theta$. (See Figure 2.)
Note that 
$$ 2z\frac{dz}{dx} = \frac{d(z^2)}{dx} = -2x(1-e^2)^{-1},$$
thus
$$ \frac{dz}{dx} = - {\tan \theta \over 1-e^2}.$$

A point on the ellipse $\vec{r} = (x,z) = (r\sin \theta, r\cos \theta)$ has
velocity $\vec{v} = \vec{r}/t$, where $t$ is the time elapsed since 
a point explosion. Constant speed (no deceleration) has been assumed.
The line-of-sight velocity of gas at this point will be
$$ v_{\rm los} = \vec{v} \cdot \hat{n} = \frac {1}{t}(x\sin i + z\cos i) $$
and the extremum, called $v_{\rm max}$,  
occurs for the point $\vec{r_*} = (x_*,z_*)$ such that
$$ {dv_{\rm los} \over dx} = {1\over t}\Bigl(\sin i + {dz \over dx}\cos i\Bigr) = 
{1 \over t}\Bigl(\sin i - {\tan \theta_* \over 1-e^2}\cos i\Bigr) = 0.$$
Thus
$$ (1-e^2)\tan i = \tan \theta_*. $$
Now $\vec{r}_*$ lies on the ellipse, so
$$ z_*^2 = a^2 -{x_*^2\over 1-e^2} = a^2 - z^2_*{\tan^2\theta_* \over 1-e^2}$$
or after some algebra,
$$ z_* = {a \over [1 + (1-e^2)\tan^2 i]^{1/2}}. $$
After some additional algebra, the desired expression is obtained:
$$ v_{\rm max} = {z_* \over t}\Bigl(\tan \theta_* \sin i + \cos i\Bigr) 
               = {a \over t}\Bigl(1 - e^2\sin^2 i\Bigr)^{1/2}
               = {1\over t}\Bigl(a^2\cos^2 i + b^2 \sin^2 i\Bigr)^{1/2}. $$
Note that in general $\theta_* \ne i$, so that the spot on the projected
image of the nova shell where $v_{\rm max}$ is observed is usually not
aligned with the central star.

\section{THE ANGLE-AVERAGED APPARENT ``RADIUS'' OF THE PROJECTED ELLIPSE}

Let the projected ellipse be described by
$$\Bigl({x \over b_\perp}\Bigr)^2 + \Bigl({y \over a_\perp}\Bigr)^2 = 1$$
where $x, y$ are now rectangular coordinates in the plane of the sky,
and $b_\perp < a_\perp$.
To streamline the notation, the projection subscript ($\perp$) is temporarily
suppressed.  Using centered polar coordinates $(r, \theta)$
such that $y = x \tan \theta$, it can be seen that
$$x^2 (a^2 + b^2\tan^2\theta) = a^2 b^2,$$
whence
$$x^2 = a^2 b^2 / (a^2 + b^2\tan^2\theta)$$
$$y^2 = a^2 b^2 \tan^2 \theta / (a^2 + b^2 \tan^2 \theta)$$
$$r^2 = x^2 + y^2 = b^2 / [\cos^2 \theta + (b/a)^2 \sin^2 \theta].$$
With the projection subscript restored, and with
$k^\prime \equiv b_\perp/a_\perp \le 1$ and $k^2 \equiv 1 - k^{\prime 2}$,
it can be seen that $r^2(\theta) = b^2_\perp / (1 - k^2\sin^2\theta)$.
The angle-averaged value of $r$ is thus
$$\bar{r} = {2 \over \pi}\int_0^{\pi/2} r(\theta)d\theta = 
{2b_\perp  \over \pi}\int_0^{\pi/2} {d\theta \over \sqrt{1-k^2\sin^2\theta} } =
{2b_\perp \over \pi} K(k) = {2b \over \pi} K(k) $$
where $K(k)$ is the complete elliptic integral of the first kind.  By
symmetry, the integration is carried out over the first quadrant
only. 

When $k^\prime \approx 1$, a useful series expansion for $K(k)$ is
(e.g. Dwight, 1961)
$$K(k) = {\pi\over 2}\Bigl(1+m\Bigr)
\Bigl[1 + {1^2\over 2^2}m^2 + {1^2 3^2\over 2^2 4^2}m^4
+ {1^2 3^2 5^2\over 2^2 4^2 6^2}m^6 + \dots\Bigr]$$
with $m \equiv  (1-k^\prime)/(1+k^\prime) = (a_\perp - b_\perp)/(a_\perp + b_\perp)$.
For $k^\prime \ge 0.4$, the series truncated after
the $m^6$ term is accurate to better than three decimal places.



\newpage

$$\vbox{
\epsfbox{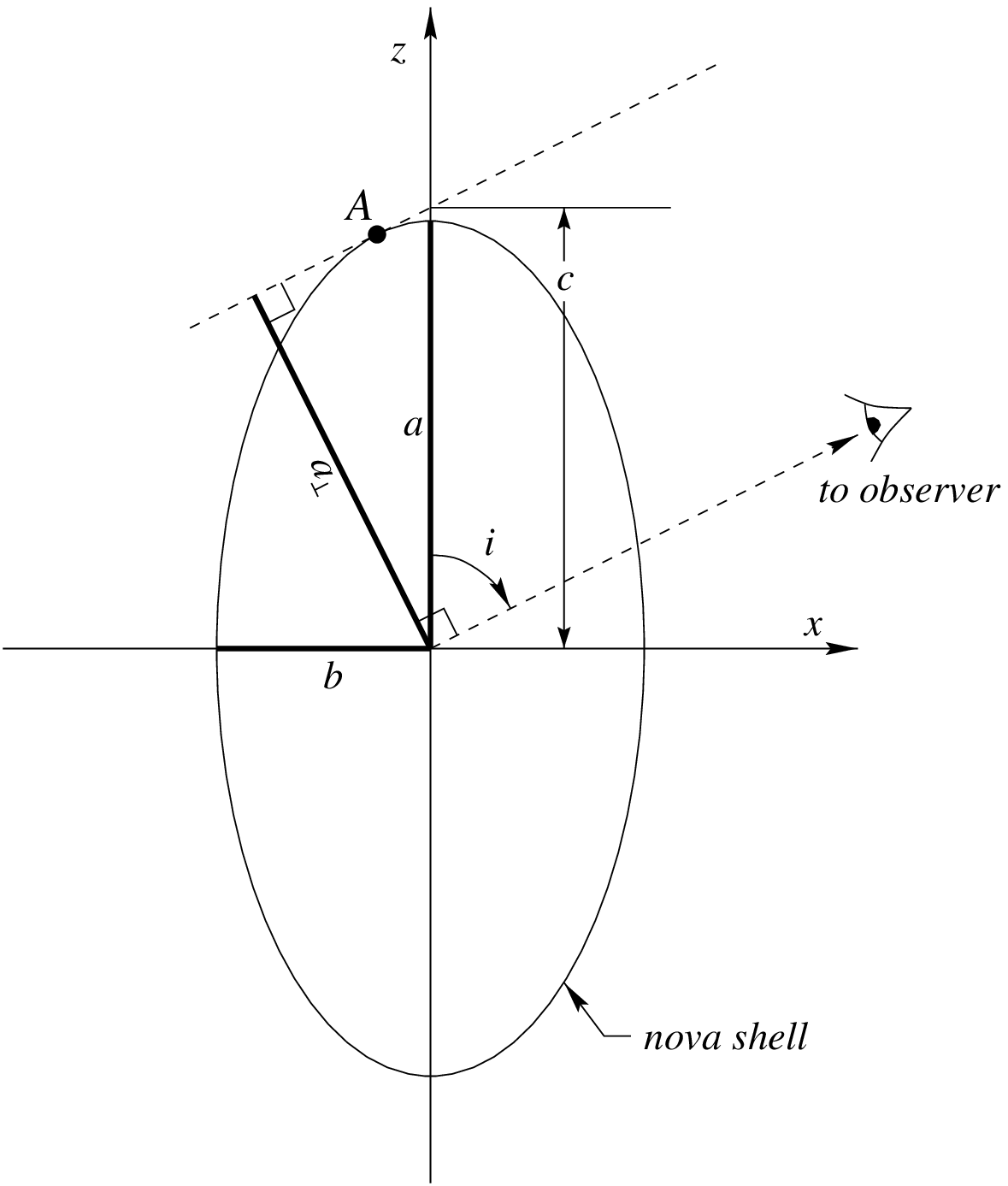}
}$$
\figcaption[fig1.eps]{Geometry in the $xz$ plane,
used in determining $a_\perp$. \label{figA1}}

\newpage

$$\vbox{
\epsfbox{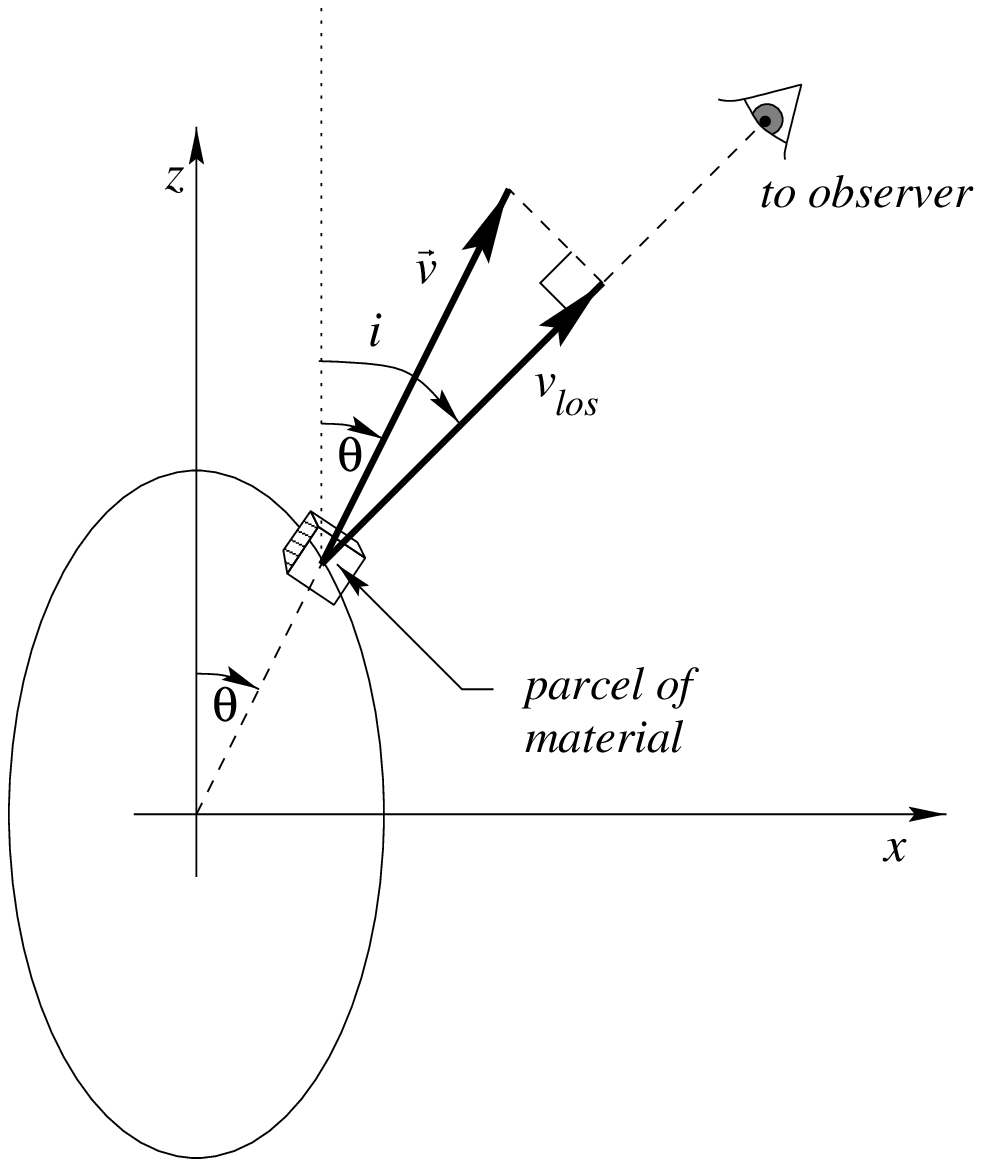}
}$$
\figcaption[fig2.eps]{Geometry for the line-of-sight
velocity of a parcel of material in a nova shell. \label{figA2}}


\begin{deluxetable}{rcccrrrrr}
\small
\tablecaption{Projection Factors and Distance Bias Factors for Prolate Ellipsoids}
\tablehead{
\colhead{$i$} & \colhead{$k^\prime$} 
& \colhead{$f_1$} & \colhead{$f_2$} & \colhead{$\hat{d}_1/d$} 
& \colhead{$\hat{d}_2/d$} & \colhead{$\hat{d}_3/d$} 
& \colhead{$\hat{d}_4/d$} & \colhead{$\hat{d}_5/d$}\\
  \colhead{(deg)}       & \colhead{($=b_\perp/a_\perp$)}
& \colhead{($=v_{\rm max}/v_0$)} & \colhead{($=a_\perp/a$)} & \colhead{ }
& \colhead{ }       & \colhead{ }       
& \colhead{ }       & \colhead{ }}
\tablecolumns{9}
\startdata
\sidehead{$b/a = f_3 = 1.00$}
0--90&1.000 & 1.000 & 1.000&  1.000 & 1.000 & 1.000 & 1.000 & 1.000\nl
\sidehead{$b/a = f_3 = 0.80$}
 0. & 1.000 & 1.000 & 0.800 & 1.250 & 1.250 & 1.250 & 1.250 & 1.250\nl
15. & 0.982 & 0.988 & 0.815 & 1.212 & 1.235 & 1.223 & 1.223 & 1.223\nl
30. & 0.936 & 0.954 & 0.854 & 1.117 & 1.192 & 1.153 & 1.154 & 1.154\nl
45. & 0.883 & 0.906 & 0.906 & 1.000 & 1.132 & 1.062 & 1.064 & 1.065\nl
60. & 0.839 & 0.854 & 0.954 & 0.896 & 1.068 & 0.974 & 0.978 & 0.980\nl
75. & 0.810 & 0.815 & 0.988 & 0.825 & 1.019 & 0.912 & 0.917 & 0.919\nl
90. & 0.800 & 0.800 & 1.000 & 0.800 & 1.000 & 0.889 & 0.894 & 0.897\nl
(averaged)  & 0.858 & 0.870 & 0.936 & 0.937 & 1.087 & 1.005 & 1.009 & 1.010\nl
\sidehead{$b/a = f_3 = 0.60$}
 0. & 1.000 & 1.000 & 0.600 & 1.667 & 1.667 & 1.667 & 1.667 & 1.667\nl
15. & 0.945 & 0.978 & 0.635 & 1.541 & 1.631 & 1.585 & 1.585 & 1.586\nl
30. & 0.832 & 0.917 & 0.721 & 1.271 & 1.528 & 1.387 & 1.393 & 1.396\nl
45. & 0.728 & 0.825 & 0.825 & 1.000 & 1.374 & 1.158 & 1.172 & 1.180\nl
60. & 0.655 & 0.721 & 0.917 & 0.787 & 1.202 & 0.951 & 0.972 & 0.983\nl
75. & 0.613 & 0.635 & 0.978 & 0.649 & 1.058 & 0.804 & 0.828 & 0.841\nl
90. & 0.600 & 0.600 & 1.000 & 0.600 & 1.000 & 0.750 & 0.775 & 0.787\nl
(averaged)  & 0.695  & 0.747   & 0.880  & 0.886 & 1.245 & 1.030 & 1.048 & 1.057\nl
\sidehead{$b/a = f_3 = 0.40$}
 0. & 1.000 & 1.000 & 0.400 & 2.500 & 2.500 & 2.500 & 2.500 & 2.500\nl
15. & 0.860 & 0.971 & 0.465 & 2.089 & 2.429 & 2.246 & 2.252 & 2.256\nl
30. & 0.658 & 0.889 & 0.608 & 1.461 & 2.222 & 1.763 & 1.802 & 1.822\nl
45. & 0.525 & 0.762 & 0.762 & 1.000 & 1.904 & 1.311 & 1.380 & 1.416\nl
60. & 0.450 & 0.608 & 0.889 & 0.684 & 1.521 & 0.944 & 1.020 & 1.061\nl
75. & 0.412 & 0.465 & 0.971 & 0.479 & 1.163 & 0.678 & 0.746 & 0.783\nl
90. & 0.400 & 0.400 & 1.000 & 0.400 & 1.000 & 0.571 & 0.632 & 0.666\nl
(averaged)  & 0.506  & 0.637   & 0.832  & 0.861 & 1.592 & 1.100 & 1.161 & 1.193\nl
\enddata
\end{deluxetable}



\end{document}